\documentclass[conference]{IEEEtran}
\IEEEoverridecommandlockouts
\usepackage{cite}
\usepackage{amsmath,amssymb,amsfonts}
\usepackage{algorithmic}
\usepackage{graphicx}
\usepackage{textcomp}
\usepackage{xcolor}
\usepackage{bm}
\usepackage{subcaption}
\def\BibTeX{{\rm B\kern-.05em{\sc i\kern-.025em b}\kern-.08em
    T\kern-.1667em\lower.7ex\hbox{E}\kern-.125emX}}

\newcommand{\figref}[1]  {Fig.~\ref{#1}}

\newcommand{\secref}[1]  {section~\ref{#1}}

\begin{document}

\title{Nonmodular architectures of cognitive systems based on active inference}

\author{\IEEEauthorblockN{Manuel Baltieri, Christopher L. Buckley}
\IEEEauthorblockA{\textit{EASY group, Sussex Neuroscience - Department of Informatics} \\
\textit{University of Sussex}\\
Brighton, UK \\
m.baltieri@sussex.ac.uk, c.l.buckley@sussex.ac.uk}
}

\maketitle

\begin{abstract}
In psychology and neuroscience it is common to describe cognitive systems as input/output devices where perceptual and motor functions are implemented in a purely feedforward, open-loop fashion. On this view, perception and action are often seen as encapsulated modules with limited interaction between them. While embodied and enactive approaches to cognitive science have challenged the idealisation of the brain as an input/output device, we argue that even the more recent attempts to model systems using closed-loop architectures still heavily rely on a strong separation between motor and perceptual functions. Previously, we have suggested that the mainstream notion of modularity strongly resonates with the separation principle of control theory. In this work we present a minimal model of a sensorimotor loop implementing an architecture based on the separation principle. We link this to popular formulations of perception and action in the cognitive sciences, and show its limitations when, for instance, external forces are not modelled by an agent. These forces can be seen as variables that an agent cannot directly control, i.e., a perturbation from the environment or an interference caused by other agents. As an alternative approach inspired by embodied cognitive science, we then propose a nonmodular architecture based on the active inference framework. We demonstrate the robustness of this architecture to unknown external inputs and show that the mechanism with which this is achieved in linear models is equivalent to integral control.
\end{abstract}


\section{Introduction}
In cognitive science it is often assumed that agents can be described as input/output systems, an idea based on traditional, computational accounts of cognition \cite{newell1972human, fodor1983modularity, hurley2001perception}. In these models, the emphasis is on internal models of the world, central processing and the sense-model-plan-act framework, often neglecting embodiment, situatedness and feedback from the environment \cite{brooks1991new}. More recent attempts, e.g., \cite{kawato1999internal, wolpert2000computational, todorov2004optimality}, have proposed closed-loop descriptions of cognitive system using internal forward/inverse models in an attempt to provide better accounts of behaviour in living organisms. However in both the input/output and the closed-loop architectures advocated by these approaches, the role of perceptual and motor processes is thought to be fundamentally modular \cite{fodor1983modularity}, i.e., these functions can be described as nearly independent, (informationally) encapsulated components with minimal interactions.

In recent years, theories of estimation and control have become increasingly popular accounts of perception \cite{knill1996perception, rao1999predictive, lee2003hierarchical} and action \cite{kawato1999internal, wolpert2000computational, todorov2004optimality} respectively. In this context, the Kalman-Bucy filter is used as a model of perception \cite{rao1999predictive, wolpert2011principles} while LQR (linear quadratic regulator) constitutes the basis of various accounts of motor control \cite{li2004iterative, stevenson2009bayesian}. In previous work \cite{baltieri2018modularity} we claimed that the idea of modularity of action and perception can be seen as an analogy of the separation principle in control theory \cite{wonham1968separation, anderson1990optimal, stengel1994optimal}. According to this principle, problems of estimation and control of a system can be solved separately and their solutions can be \emph{optimally} combined under a set of assumptions. Following this, one can sequentially combine a Kalman-Bucy filter and LQR to create the LQG (linear quadratic Gaussian) architecture, used as a general methodology for several models of sensorimotor loops, e.g., \cite{wolpert2000computational, todorov2002optimal, stevenson2009bayesian, yeo2016optimal}. The ``classical sandwich'' \cite{hurley2001perception} of cognitive science thus survives, we claim, even in the forward/inverse models formulation of perception and motor control.

The fields of embodied and enactive cognitive science on the other hand emphasise the deep integration of perception and action, seen as fundamentally intertwined \cite{clark1998being, wilson2002six, di2017sensorimotor}. In \cite{baltieri2018modularity} we proposed to use a framework based on the formulation of perception and action as estimation and control while not implementing the conditions for the separation principle, i.e., active inference. Active inference is a process theory based on the free energy principle \cite{Friston2010nature} describing cognitive functions (perception and action, but also learning and attention) as processes of minimisation of sensory surprisal \cite{Friston2010nature, buckley2017free}. More precisely, since this quantity is not directly accessible by an agent, it is thought that the variational free energy (an upper bound to sensory surprisal) is minimised in its place. In active inference, perceptual and motor processes are often described as entangled and inseparable \cite{friston2011optimal, wiese2016action, pezzulo2017model} providing thus a new possible methodology combining estimation and control following embodied/enactive theories of the mind. We previously presented a conceptual account of active inference and its role for nonmodular architectures of cognitive systems \cite{baltieri2018modularity}. Here we introduce a minimal agent model highlighting the different implementations (LQG vs. active inference) especially in presence of unknown external stimuli affecting an agent's observations.

\section{LQG and the separation principle}
The framework provided by LQG control and based on the separation principle linearly combines two processes of 1) estimation or inference of hidden properties of the environment and 2) control or regulations of variables of interest. The estimation of hidden variables is based on the presence of a Kalman (for discrete time systems) or Kalman-Bucy (for continuous time systems) filter, while the control of the desired variables on LQR \cite{wonham1968separation, anderson1990optimal, stengel1994optimal}. In particular, this combination is provably optimal according to a set of assumptions:
\begin{enumerate}
  \item the estimator is implemented through a state-space model where only linear process dynamics and observation laws describe the environment and its latent states
  \item uncertainty or noise in both dynamics and observations are represented by \emph{white}, zero-mean Gaussian variables
  \item the properties of these random variables, in particular their (co)variance matrices, are known
  \item the performance of the regulator can be evaluated using a quadratic cost function
  \item all the inputs/forces applied to the agent are known, e.g., external disturbances and internal signals such as motor actions.
\end{enumerate}

Following the separation principle, the LQG controller produces optimal estimation and optimal control for linear systems, sequentially combining two separate sub-systems, a Kalman-Bucy filter and LQR, in an optimal (i.e., minimum-variance) way \cite{anderson1990optimal, stengel1994optimal}. The Kalman-Bucy filter provides the optimal state-estimate of a signal and the LQR controller uses such estimate (i.e., the mean) to implement the optimal deterministic controller: LQG control makes use of the estimated mean and feeds it into an LQR controller.

A general linear system to be regulated in the presence of noise on the observed state is described by:
\begin{align}
    d \bm{x} = A \bm{x} \: dt + B \bm{a} \: dt + d \bm{w} \quad \quad \quad
    \bm{y} = C \bm{x} + d\bm{z}
    \label{eq:LQGSSM}
\end{align}
where all the variables and parameters are the same as previously defined for Kalman-Bucy filters and LQR. Using the separation principle, it can then be shown that minimising the expected value of the cost-to-go is equivalent to minimising the cost-to-go for the expected (estimated) state \cite{stengel1994optimal}
\begin{align}
    c(\bm{x}, \bm{a}) = c(\hat{\bm{x}}, \bm{a}) = \frac{1}{2} \hat{\bm{x}}^T Q \hat{\bm{x}} + \frac{1}{2} \bm{a}^T R \bm{a}
    \label{eq:costrateLQG}
\end{align}
where we replaced states $\bm{x}$ with their estimates $\hat{\bm{x}}$, meaning that the optimal control can be computed using only the state estimate (i.e., the mean) $\hat{\bm{x}}$ rather than $\bm{x}$. The combined problem of estimation and control in LQG terms is then implemented by the following system combining Kalman-Bucy filter and LQR equations:
\begin{align}
    \dot{\hat{\bm{x}}} = & A \hat{\bm{x}} + B \bm{a} + K (y - C \hat{\bm{x}}) \IEEEnonumber \\
    \bm{a} = & - L \hat{\bm{x}} \IEEEnonumber \\
    K = & P H^T (\Sigma_z)^{-1} \IEEEnonumber \\
    L = & R^{-1} B^T V \IEEEnonumber \\
    \dot{P} = & \Sigma_w + A P + P A^T - K (\Sigma_z) K^T \IEEEnonumber \\
    - \dot{V} = & Q + A^T V + V A - L^T R L.
    \label{eq:LQGEstimationControl}
\end{align}

\section{Active inference}
\label{sec:AI}
Active inference is a process theory proposed to explain brain functioning and other functions of living systems based on Bayesian inference and optimal control theory \cite{Friston2010biocyb, Friston2010nature, buckley2017free}. In this section we establish its relations to the LQG architecture, starting by building an active inference version of the regulation of a linear multivariate system, and highlighting differences, limitations and possible extensions proposed for the control problem. As with LQG control, we build an estimator of the hidden states $\bm{x}$. In this case however, we will give a variational account of the estimator in generalised coordinates of motion that generalises the MLE/MAP derivation of Kalman-Bucy filters \cite{chen2003bayesian} using Variational Bayes with a Laplace approximation \cite{Friston2008a, buckley2017free}. We start by defining a generative model for an agent capturing the dynamics of the system to control and how these relate to observations and represented in a \emph{generalised} state-space form \cite{Friston2008c, buckley2017free}:
\begin{align}
    \bm{x}' = \hat{A} \bm{x}' + \hat{B} \bm{v} + \bm{w} \quad \quad \quad
    \bm{y} = \hat{C} \bm{x} + \bm{z}
\end{align}
where the hat over the matrices simply represents the fact that the matrices used in the generative model don't necessarily mirror their counterparts describing the world dynamics (\cite{baltieri2017active, brown2013active}), as shown in our model later. The main difference with respect to LQG however is that LQG explicitly mirrors (by construction in the linear case) the dynamics of the observed system, thus including knowledge of inputs $\bm{a}$. On the other hand, in active inference this vector is not explicitly modelled by an agent, assuming that such information is not available to a system, in accordance with evidence in motor neuroscience suggesting the lack of knowledge of self-produced controls (i.e., efference copy) \cite{feldman2009new, friston2011optimal, feldman2016active}. It is in fact proposed that a deeper duality of estimation and control exists whereby, in the simplest case (i.e., a purely reflexive account), actions are simply responses to the presence of prediction errors at the proprioceptive level, irrespectively of the cause of sensations (self-generated or external forces) \cite{friston2011optimal, brown2013active}. The vector $\bm{v}$ in the generative model encodes instead external or exogenous inputs in a state-space models context or, from a Bayesian perspective, priors or ``desired'' outcomes generated by higher layers in hierarchical (Bayesian) implementations \cite{lee2003hierarchical, Friston2008c}. In this light, priors can be used to effectively bias the estimator to ``infer'' desired rather than observed states, with a controller instantiating actions on the world to fulfil the ``observed'' (= desired) states of an agent. Variables $\bm{z}, \bm{w}$ model the real noise in the environment making, however, use of the definition of state space models in generalised coordinates of motion \cite{Friston2008a, Friston2008c}, where $z, w$ are treated as analytical noise with non-zero autocorrelation, generalising the definition of Wiener processes with Markov property.

This state-space model can then be written down in a probabilistic form, mapping the measurements equation to a likelihood $P(\bm{y} | \bm{\hat{x}})$ (no direct influence of inputs on observations), and a the dynamics to a prior $P(\bm{\hat{x}}, \bm{v})$ \cite{Friston2008c, buckley2017free, baltieri2017active}. The two multivariate Gaussian probabilities densities can then be combined and used in the general formulation of Laplace encoded variational free energy defined in \cite{Friston2008a, buckley2017free} (without constants):
\begin{align}
  F \approx - \ln P(\bm{y}, \bm{x}, \bm{v}) \Bigr \rvert_{\bm{x} = \bm{\mu_x}, \bm{v} = \bm{\mu_v}}
  \label{eq:freeEnergyLaplace}
\end{align}
the free energy for a generic linear multivariate system becomes then:
\begin{align}
    F & \approx \frac{1}{2} \bigg[ \Big( \bm{y} - \hat{C} \bm{\mu_x} \Big)^T \Pi_z \Big( \bm{y} - \hat{C} \bm{\mu_x} \Big) + \IEEEnonumber \\
    & + \Big (\bm{\mu_x'} - \hat{A} \bm{\hat{\mu}_x} - \hat{B} \bm{\mu_v} \Big)^T \Pi_w \Big (\bm{\mu_x'} - \hat{A} \bm{\mu_x} - \hat{B} \bm{\mu_v} \Big) + \IEEEnonumber \\
    & - \ln \big| \Pi_z \big| - \ln \big| \Pi_w \big| + (m + n) \ln 2 \pi \bigg]
    \label{eq:freeEnergyMultivariate}
\end{align}
where we explicitly replaced $\bm{x}, \bm{v}$ with their expectations $\bm{\mu_x}, \bm{\mu_v}$ since under the Laplace assumption this represents the best estimate of $\bm{x}, \bm{v}$ (i.e., covariances of the approximate, variational density can be recovered analytically \cite{Friston2008a, buckley2017free}). Variables $m, n$ represent the length of vectors $\bm{y}$ and $\bm{x}$ respectively. Expectations $\bm{\mu_x}$ play the same role of estimates $\bm{\hat{x}}$ in LQG, we simply decided to use a notation consistent with some of our previous work \cite{buckley2017free, baltieri2017active, baltieri2018propabilistic}. We also defined precision matrices $\Pi_z, \Pi_w$ as the inverse of covariance matrices $\Sigma_z, \Sigma_w$ and used $|\cdot|$ to define the determinant of a matrix. It is important to highlight that, in general, the covariance matrices used in the generative model can be different from the ones used to describe the environment or generative process \cite{brown2013active, baltieri2017active}. To simplify the already heavy notation we will however represent them in the same way.

The recognition dynamics, encoding perception and action in a system minimising free energy \cite{Friston2008a, buckley2017free} and equivalent to estimation and control functions respectively, are implemented in standard active inference formulations as a gradient descent scheme minimising the free energy with respect to the variables $\bm{\mu_{x}}$ for perception/estimation:
\begin{align}
    \bm{\dot{\mu}_x} & = D \bm{\mu_x} - \frac{\partial F}{\partial \bm{\mu_x}} = \bm{\mu_x'} + \hat{C}^T \Pi_z \Big( \bm{y} - \hat{C} \bm{\mu_x} \Big) + \IEEEnonumber \\
    & + \hat{A}^T \Pi_w \Big (\bm{\mu_x'} - \hat{A} \bm{\mu_x} - \hat{B} \bm{\mu_v} \Big) \IEEEnonumber \\
    \bm{\dot{\mu}_x'} & = D \bm{\mu_x'} - \frac{\partial F}{\partial \bm{\mu_x'}} = \bm{\mu_x''} - \Pi_w \Big (\bm{\mu_x'} - \hat{A} \bm{\mu_x} - \hat{B} \bm{\mu_v} \Big)
    \label{eq:perceptionActiveInference}
\end{align}
and actions $\bm{a}$ for action/control, assuming only that actions have an effect on observations $\bm{y}$ \cite{Friston2010biocyb}:
\begin{align} 
    \bm{\dot{a}} & = - \frac{\partial F}{\partial \bm{a}} = - \frac{\partial F}{\partial \bm{y}} \frac{\partial \bm{y}}{\partial \bm{a}} = - \frac{\partial \bm{y}}{\partial \bm{a}}^T \Pi_{z} \Big( \bm{y} - \hat{C} \bm{\mu_{x}} \Big).
    \label{eq:actionActiveInference}
\end{align}
The estimation expressed in \eqref{eq:perceptionActiveInference} prescribes a generalisation of Kalman-Bucy filters to trajectories with arbitrary embedding orders where random variables are not treated as Markov processes \cite{Friston2008a}. In \eqref{eq:perceptionActiveInference}, we also include an extra term $D \bm{\mu_{x}}$ that represents the ``mode of the motion'' (also the mean for Gaussian variables) for the minimisation in generalised coordinates of motion \cite{Friston2008c, buckley2017free}, with $D$ as a differential operator shifting the order of motion, i.e., $D \bm{\mu_{x}} = \bm{\mu_{x}}'$. More intuitively, since we are now minimising the components of a generalised state representing a trajectory rather than a static variable, variables are in a moving framework of reference where the minimisation is achieved for $\bm{\dot{\mu}_x} = \bm{\mu'_x}$ rather than $\bm{\dot{\mu}_x} = \bm{0}$. Action as expressed in \eqref{eq:actionActiveInference} may appear similar to the traditional LQR/LQG form, but is fundamentally different since it depends explicitly on observations $\bm{y}$ rather than estimated hidden states $\bm{\mu_x}$.

\section{The model}
The double integrator is a canonical example used in control theory and represents one of the most fundamental problems in optimal control, modelling single degree of freedom motion of different physical systems \cite{rao2001naive, aastrom2010feedback}. In the case presented here, this could be thought of as a block on frictionless surface. In motor neuroscience, this is the simplest model of single-joint movement \cite{gottlieb1993computational} and can, in some cases, be easily generalised to multiple degrees of freedom \cite{Friston2010biocyb}. The standard double integrator is usually described as a deterministic system. The control policy is thus defined using a feedback law applied directly to the \emph{known} dynamics, as the full state of the system is measured with no uncertainty \cite{rao2001naive}. For the purposes of this work, where uncertainty and noise are crucial components, we will introduce process and measurement noise into the system, making the estimation of hidden states necessary. This will then allow us to compare LQG and active inference in one of the simplest possible examples in the control theory literature with direct applications to the study of motor systems and behaviour \footnote{The code is available at https://github.com/mbaltieri/doubleIntegrator}. The double integrator is described by the following state-space model:
\begin{equation}
  \begin{aligned}
    \dot{\bm{x}} = A \bm{x} + B \bm{a} + \bm{w}
  \end{aligned}
  \quad
  \begin{aligned}
    \bm{y} = C \bm{x} + \bm{z}
  \end{aligned}
  \label{eq:doubleIntegratorSSM}
\end{equation}
where matrices $A, B, C$ are defined as:
\begin{align}
    A = 
    \begin{bmatrix} 
      0 & 1 \\
      0 & 0
    \end{bmatrix}\IEEEnonumber
  	\quad
    B = 
    \begin{bmatrix} 
      0 & 0\\
      0 & 1
    \end{bmatrix}\IEEEnonumber
    \quad
    C = 
    \begin{bmatrix} 
      1 & 0 \\
      0 & 1
  	\end{bmatrix}\IEEEnonumber
    \label{eq:doubleIntegratorMatrices}
\end{align}
and covariance matrices $\Sigma_z, \Sigma_w$ as:
\begin{align}
    \Sigma_z = 
    \begin{bmatrix} 
      \exp(0) & 0 \\
      0 & \exp(0)
    \end{bmatrix}
  	\quad
    \Sigma_w = 
    \begin{bmatrix} 
      0 & 0 \\
      0 & \exp(-1)
    \end{bmatrix}\IEEEnonumber
\end{align}

\begin{figure}[ht!]
  \centering
  \includegraphics[width=.7\linewidth]{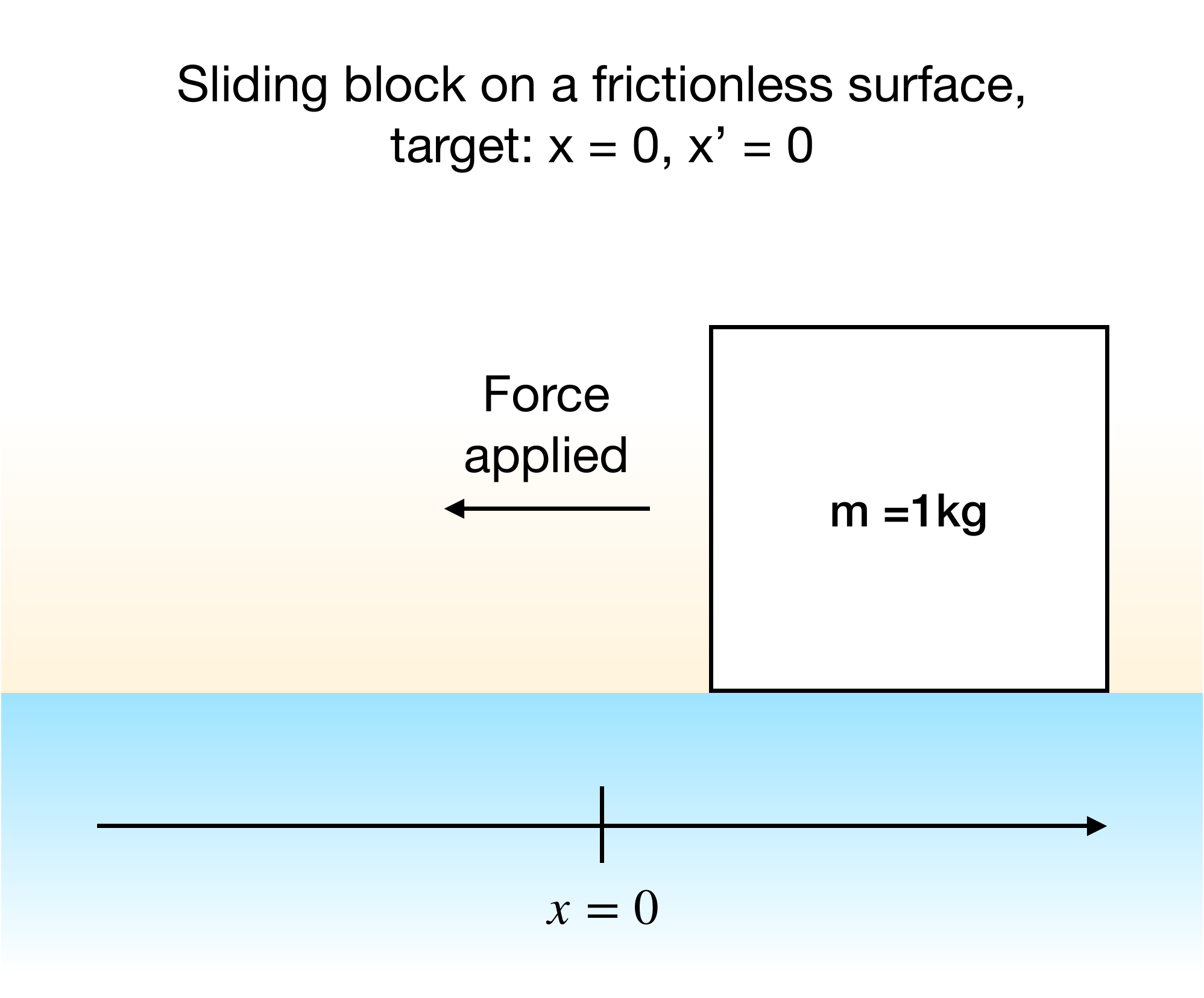}
  \caption{\textbf{The generative process, a double integrator.} The double integrator models the motion of a system with a single degree of freedom, corresponding to a block of mass=1kg placed on a surface with no friction. The block is initialised at a random position with a random velocity and needs to stop, $x'=0$, at position $x=0$.}
  \label{fig:DoubleIntegratorGP}
\end{figure}

\subsection{The LQG solution to the double integrator}
For LQG we implement \eqref{eq:LQGEstimationControl} using the same matrices $A, B, C, \Sigma_z, \Sigma_w$ specified above and furthermore define:
\begin{equation}
  \begin{aligned}
    Q = 
    \begin{bmatrix} 
      1 & 0 \\
      0 & 1
    \end{bmatrix}
  	\quad
    R = 
    \begin{bmatrix} 
      4 & 0 \\
      0 & 4
    \end{bmatrix}
  \end{aligned}
  \label{eq:doubleIntegratorLQRWeights}
\end{equation}
with no specific optimisation of these parameters since it is beyond the scope of this work. For further analysis see for instance \cite{rao2001naive}.
\begin{figure}[ht!]
	\begin{subfigure}{.5\textwidth}
		\centering
		\includegraphics[width=\linewidth]{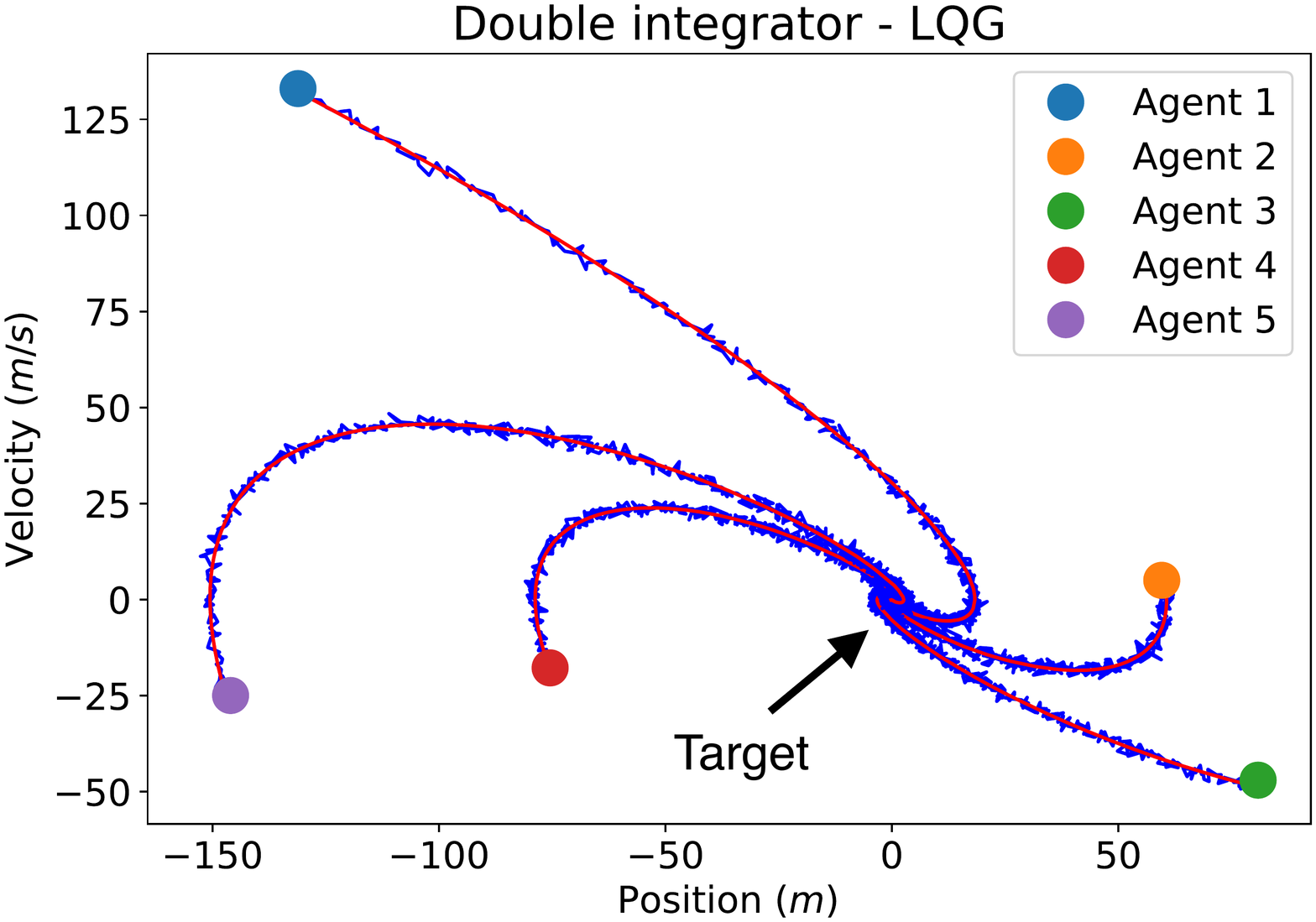}
		\caption{}
		\label{fig:DoubleIntegratorLQGAgents}
	\end{subfigure}
	\begin{subfigure}{.5\textwidth}
		\centering
		\includegraphics[width=\linewidth]{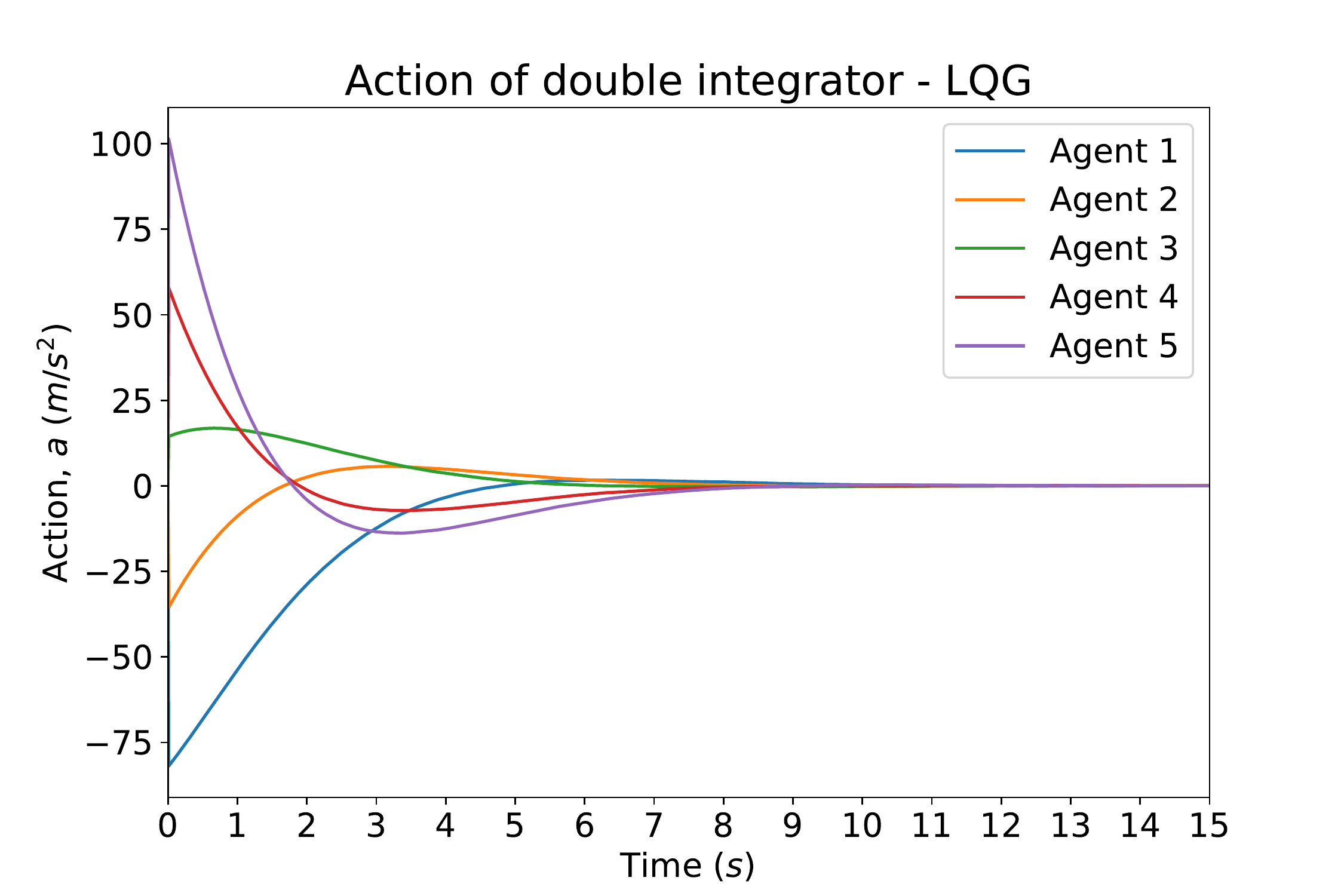}
		\caption{}
		\label{fig:DoubleIntegratorLQGAction}
	\end{subfigure}
  \caption{\textbf{The double integrator solved using LQG.} (a) Five examples with different initial conditions showing in blue the observed trajectories of different blocks in the phase-space and in red the agent's estimates of the same trajectories. (b) Actions taken by the five agents.}
  \label{fig:DoubleIntegratorLQG}
\end{figure}
\begin{figure}[ht!]
  \begin{subfigure}{.5\textwidth}
    \centering
    \includegraphics[width=\linewidth]{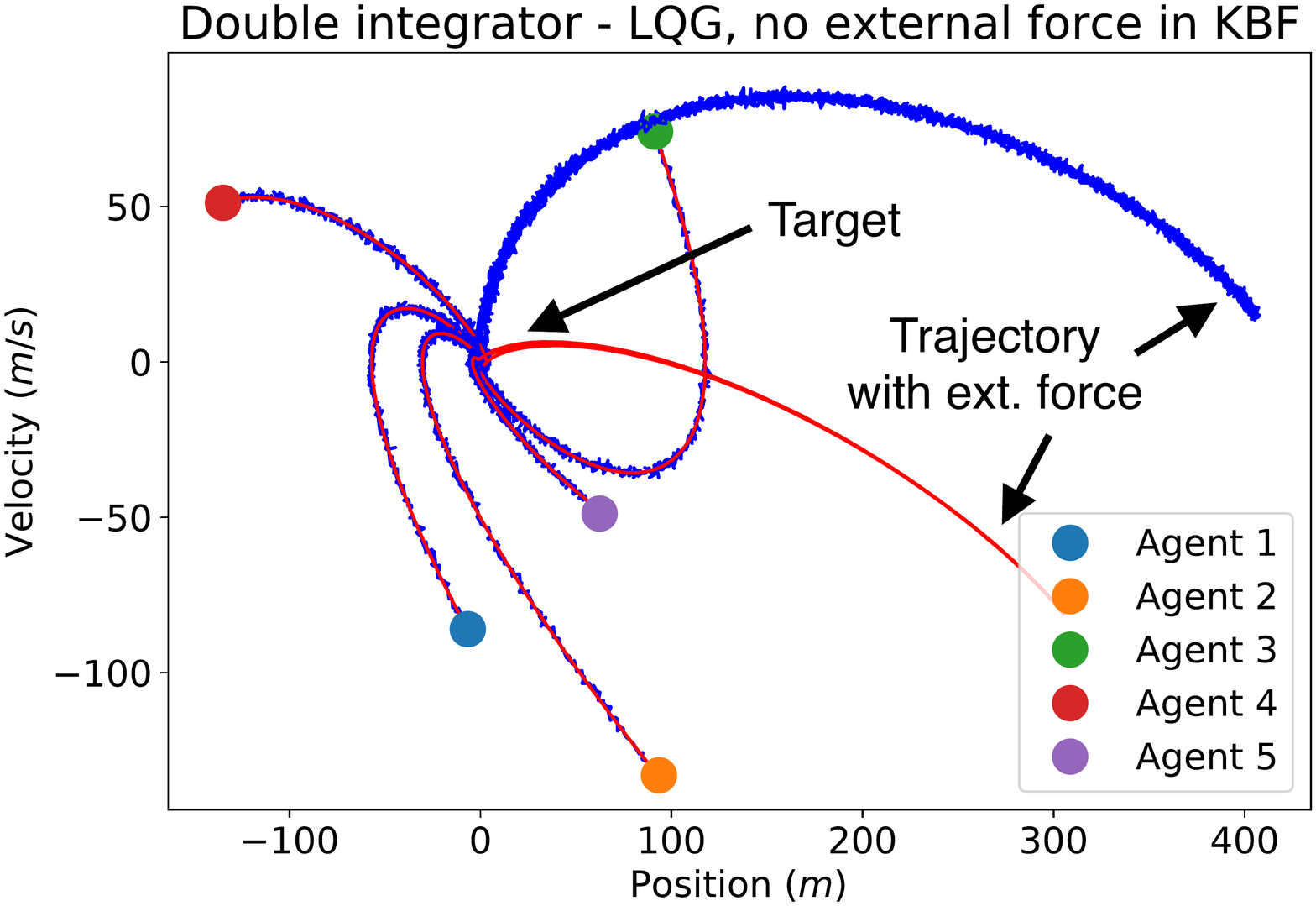}
    \caption{}
    \label{fig:DoubleIntegratorLQGNoExternalForceAgents}
  \end{subfigure}
  \begin{subfigure}{.5\textwidth}
    \centering
    \includegraphics[width=\linewidth]{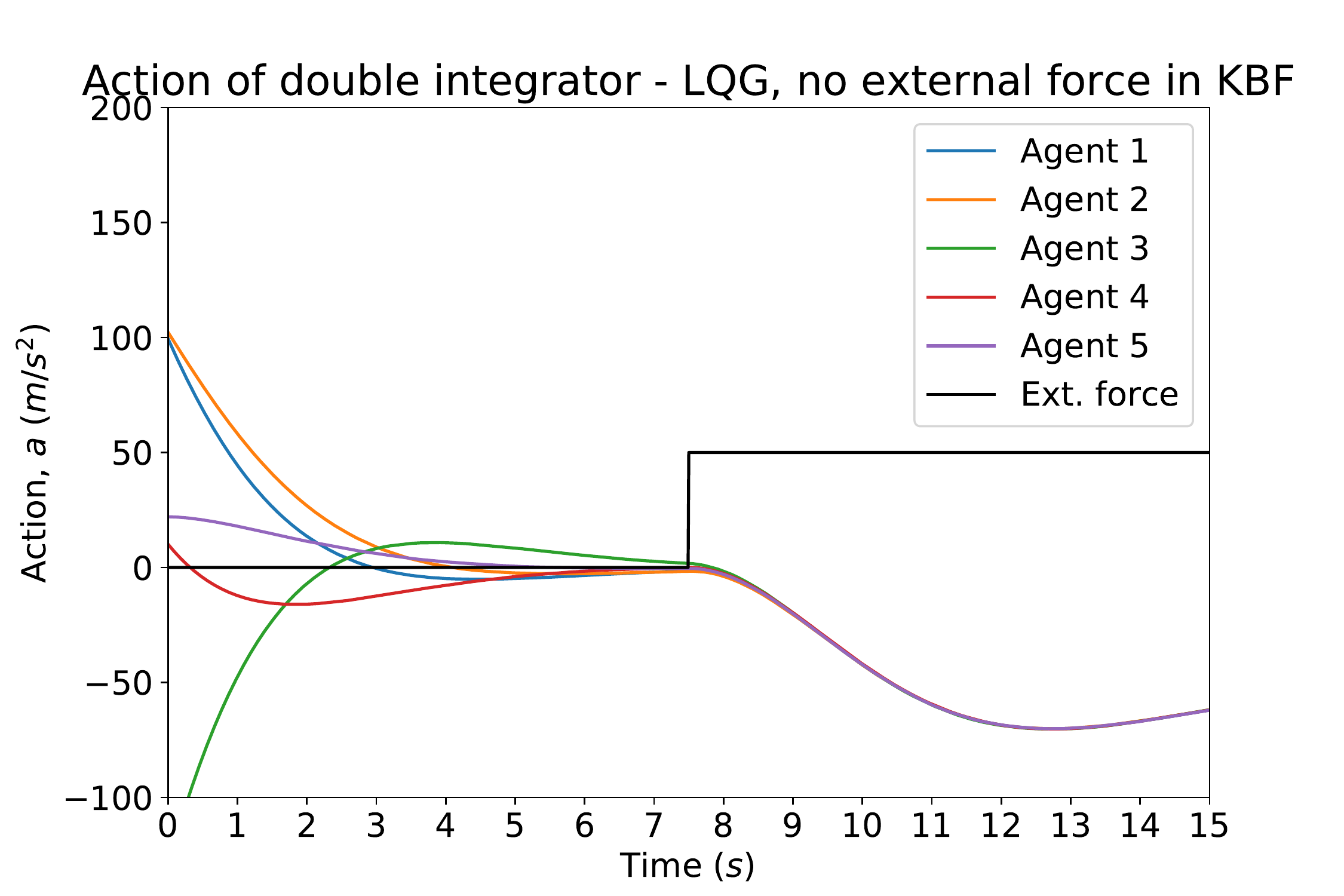}
    \caption{}
    \label{fig:DoubleIntegratorLQGNoExternalForceAction}
  \end{subfigure}
  \caption{\textbf{The double integrator solved using LQG.} (a) Five examples with different initial conditions showing in blue the observed trajectories of different blocks in the phase-space and in red the agent's estimates of the same trajectories. (b) Actions taken by the five agents after an external force is introduced (black line).}
  \label{fig:DoubleIntegratorLQGNoExternalForce}
\end{figure}
As we can see in \figref{fig:DoubleIntegratorLQGAgents}, the block is effectively driven to the desired position $x=0$ and velocity $x'=0$ from a set of 5 randomly initialised conditions (position and velocity are sampled from zero-mean Gaussian distributions, sd=300). In \figref{fig:DoubleIntegratorLQGAction} we then show the actions over time of the same 5 example agents, all converging to zero since the agents effectively reach their desired target. The main feature of LQG, and from which active inference will depart, is the reliability of estimates of both position and velocity (the red line in the phase space), using a Kalman-Bucy filter. In LQG, accurate estimates are necessary to then enact the LQR component implementing a negative feedback mechanism based on estimates $\bm{\hat{x}}$ rather than true hidden states $\bm{x}$. In \figref{fig:DoubleIntegratorLQGNoExternalForce} we introduced a new external force not modelled by the agents, equivalent to a disturbance from the environment (black line in \figref{fig:DoubleIntegratorLQGNoExternalForceAction}). \figref{fig:DoubleIntegratorLQGNoExternalForceAgents} then shows that the agents are incapable of regulating their position/velocity against this unknown input (blue lines), after an initial convergence towards the desired state, they in fact move away from it when the unexpected force is introduced. Furthermore, these agents are incapable of correctly inferring their trajectories, providing inaccurate estimates of their sensed variables (red lines). In \figref{fig:DoubleIntegratorLQGNoExternalForceAction} we see that all of these agents attempt to counteract the effects of unexpected stimuli (they minimise their velocity after the force is introduced), however the lack of an appropriate mechanism to track their position correctly (e.g., integral action) pushes them away from the target.

\subsection{The double integrator with active inference}
To solve the same control problem, active inference relies on the generation of predictions of proprioceptive sensations (position, velocity as in LQG, and also acceleration in this case), followed by the implementation of actions in the world via (trivial) reflex arcs. The proprioceptive modality is essentially treated as other inputs (vision, audition, etc.) and estimates/predictions are generated using the same generative model taking advantage of incoming proprioceptive sensations. This produces a considerably different control system, with state estimates and actions now created by the same model, making it hard to clearly separate processes of perception and action. The copy of motor control signals (cf. efference copy \cite{von1950reafferenzprinzip}), necessary in standard LQG settings to meet the observability constraints of Kalman-Bucy filters \cite{anderson1990optimal, stengel1994optimal} is not included in this formulation, as explained in \secref{sec:AI}. Active inference postulates in fact that direct representations of the causes or actions $\bm{a}$ of self-generated sensations need not be discounted during the prediction of new incoming sensory inputs. This could be seen as a limitation of active inference, but in general this speaks to the robustness of this approach in face of unknown inputs (i.e., motor actions produced by an agent or exogenous forces from the environment), see \cite{baltieri2018propabilistic}. In this framework, inputs can also be estimated using an appropriate generative model of the world dynamics \cite{Friston2008a}, a feature thought to be fundamental in biological systems \cite{sontag2003adaptation}. Simple and effective approximations are also possible, for example with integral control, thought to be the most basic heuristic dealing with the problem of uncertain inputs in biological systems down to the unicellular level \cite{yi2000robust, sontag2003adaptation} and already shown to be consistent with formulations of active inference \cite{baltieri2018propabilistic}.
\begin{figure}[ht!]
  \centering
  \includegraphics[width=.9\linewidth]{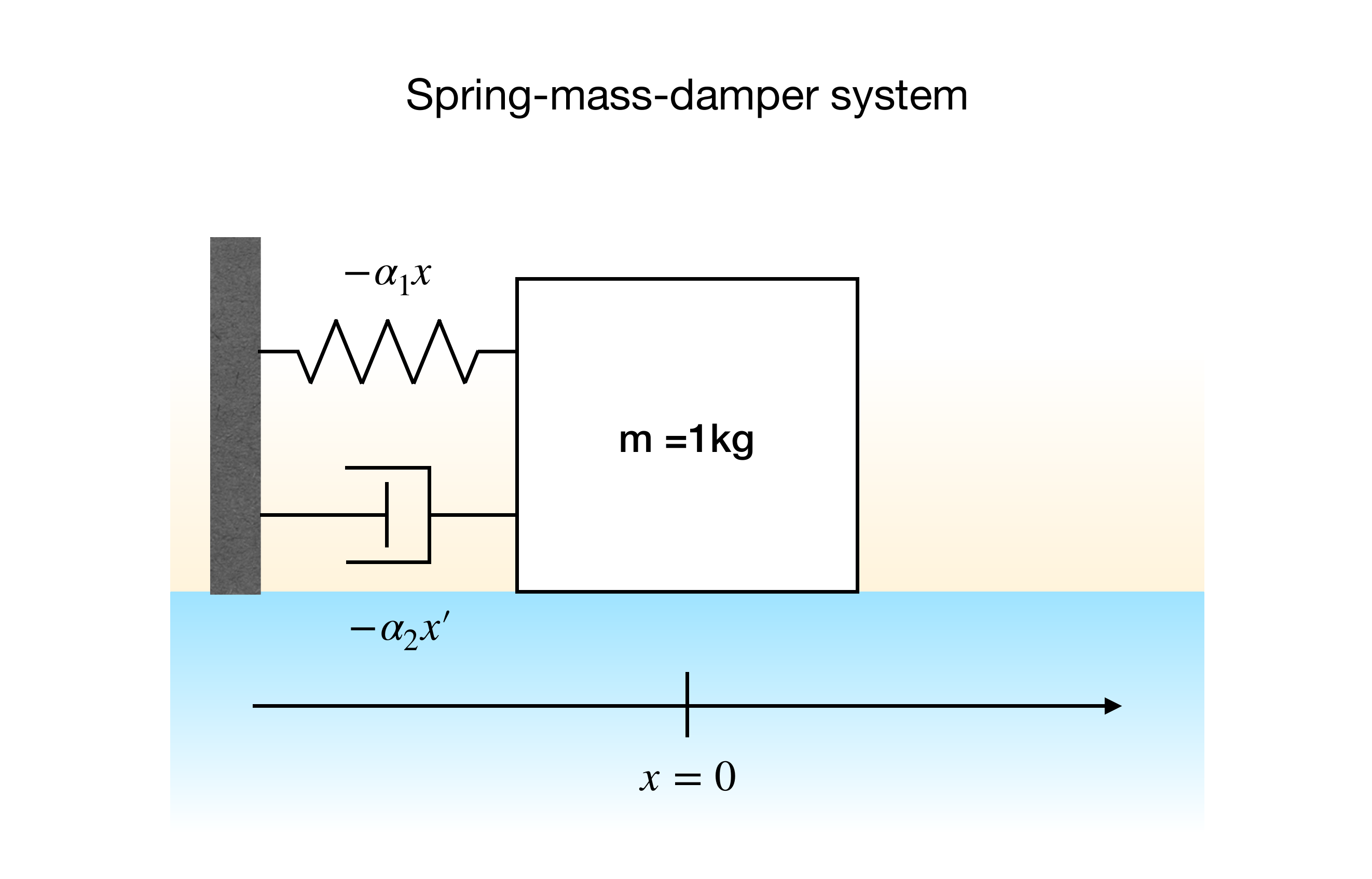}
  \caption{\textbf{The generative model.} To implement the regulation of position and velocity, the agent implements a model whereby an imaginary spring pulls the block back to the origin ($x=0$) while an imaginary damper slows it down ($x' = 0$).}
  \label{fig:DoubleIntegratorGM}
\end{figure}

To derive an active inference solution to the double integrator, we start by defining a generative model for the agent, i.e., the block:
\begin{equation}
  \begin{aligned}
    \bm{x}' = \hat{A} \bm{x} + \hat{B} \bm{v} + \bm{w}
  \end{aligned}
  \quad
  \begin{aligned}
    \bm{y} = \hat{C} \bm{x} + \bm{z}
  \end{aligned}
  \label{eq:doubleIntegratorSSMGenerativeModel}
\end{equation}
where matrix $\hat{A}$ is:
\begin{align}
    \hat{A} = 
    \begin{bmatrix} 
      0 & 1 & 0 \\
      -\alpha_1 & -\alpha_2 & 0 \\
      0 & 0 & 0
    \end{bmatrix}\IEEEnonumber
    \label{eq:doubleIntegratorMatricesActiveInference}
\end{align}
while $\hat{B}$ is diagonal with $\exp (1)$ values, $\hat{C}$ is zero everywhere but in $C_{2,2}$ where the motor action is applied (with a value of 1) and covariance matrices $\Sigma_z, \Sigma_w$ are also diagonal with, respectively, $\exp (1)$ and $\exp (8)$ values on the main diagonals. The agent implements beliefs of a world where it is pulled back to the desired state $x=x'=0$ by an imaginary spring and slows down thanks to an imaginary piston-like damper, ``designed'' (in this case by us, but more in general one could imagine evolutionary processes for biological system \cite{Friston2010biocyb}) to favour normative behaviour.

Following \eqref{eq:freeEnergyMultivariate}, the variational free energy for our controller is then described by:
\begin{align}
  F & \approx \frac{1}{2} \bigg[ \pi_{z} (y - \mu_x)^2 + \pi_{z'} (y' - \mu_x')^2 + \pi_{z''} (y'' - \mu_x'')^2 + \IEEEnonumber \\
  & + \pi_{w'} (\mu_x'' - \mu_v)^2 - \ln (\pi_{z} \pi_{z'} \pi_{z''} \pi_{w'}) + (3+2)\ln 2 \pi \bigg]
\end{align}
where precisions $\pi$ are taken from the diagonals of precision matrices $\Pi_z, \Pi_w$ (inverse covariances matrices $\Sigma_z, \Sigma_w$ defined in the generative model). After explicitly writing out the equations derived from the matrix formulation in \eqref{eq:perceptionActiveInference}, we get the following formulation of perceptual inference:
\begin{align}
    \dot{\mu}_x = & \mu_x' + \pi_{z} (y - \mu_x) + \pi_w (\mu_x' + \alpha \mu_x - \beta \mu_v) \IEEEnonumber \\
    \dot{\mu}_x' = & \mu_x'' + \pi_{z'} (y' - \mu_x') + \pi_{w'} (\mu_x'' + \alpha \mu_x' - \beta \mu_v') \IEEEnonumber\\
    \dot{\mu}_x'' = & \mu_x''' + \pi_{z''} (y'' - \mu_x'') + \pi_{w''} (\mu_x'' + \alpha \mu_x' - \beta \mu_v') 
    \label{eq:ActiveInferenceEstimation}
\end{align}
and
\begin{align}
    \dot{\mu}_x' = & - \pi_w (\mu_x' + \alpha \mu_x - \beta \mu_v) \IEEEnonumber \\
    \dot{\mu}_x'' = & - \pi_{w'} (\mu_x'' + \alpha \mu_x' - \beta \mu_v')
\end{align}
showing the lack of the Kalman gain $K$ and an important difference derived from its absence: if $K$ is non-diagonal as in this case (one can simply verify this claim with standard functions solving continuous Riccati equations, as in the provided code), both orders of motion are present in the optimal filter problem in \eqref{eq:LQGEstimationControl}, but only one appears in \eqref{eq:ActiveInferenceEstimation} since the precision matrices are assumed to be diagonal in our formulation. More in general, in active inference the Kalman gain $K$ matrix is replaced by learning rates such as in this work or \cite{baltieri2017active}, or by clever implementations that allow for adaptive update schemes with varying integration steps as in \cite{Friston2008a}.

The action component is, however, the one most significantly different, starting from the assumption that direct knowledge of motor signals is not available and thus not modelled in the generative model (motor commands $\bm{a}$ are replaced by inputs $\bm{v}$ acting as priors). This entails a new approach to the problem, with active inference suggesting that the only information needed comes from observations $\bm{y}$, see \eqref{eq:actionActiveInference}. On this account, action reduces to
\begin{equation}
  	\bm{\dot{a}} = - \bigg( \frac{\partial \bm{y}'}{\partial a} \bigg)^T \Pi_{z} (\bm{y} - \hat{C} \bm{\mu}_x)
  	\label{eq:ActiveInferenceControlMatrixVersion}
\end{equation}
and with the assumption that 
\begin{equation*}
    \frac{\partial \bm{y}}{\partial a} = 
    \begin{bmatrix} 
        1 & 1 & 1
    \end{bmatrix}^T
\end{equation*}
the explicit, scalar version of action becomes
\begin{equation}
    \dot{a} = - \pi_{z} (y - \mu_x) - \pi_{z'} (y' - \mu_x') - \pi_{z''} (y'' - \mu_x''),
\end{equation}
replacing the LQR component in \eqref{eq:LQGEstimationControl}. This type of control is equivalent to a PID controller, and is the ``optimal'' linear solution when knowledge of inputs $\bm{a}$ is not available in the generative model \cite{baltieri2018propabilistic}. As in the case of filtering, the feedback gain $L$ is missing in the active inference formulation, once again replaced by learning rates of the gradient descent or by other approximations. 
\begin{figure}[ht!]
	\begin{subfigure}{.5\textwidth}
		\centering
		\includegraphics[width=\linewidth]{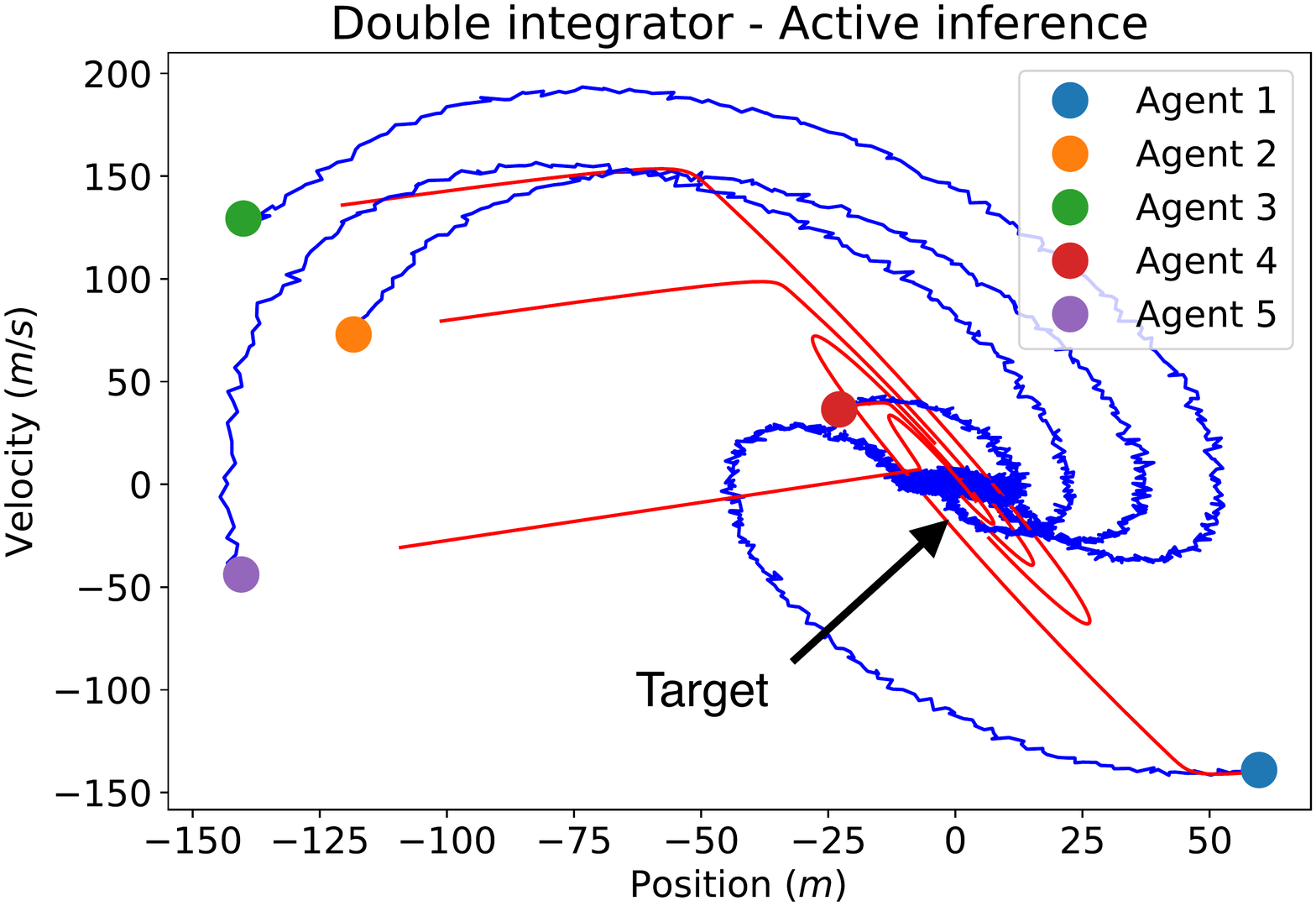}
		\caption{}
		\label{fig:DoubleIntegratorActiveInferenceAgents}
	\end{subfigure}
	\begin{subfigure}{.5\textwidth}
		\centering
		\includegraphics[width=\linewidth]{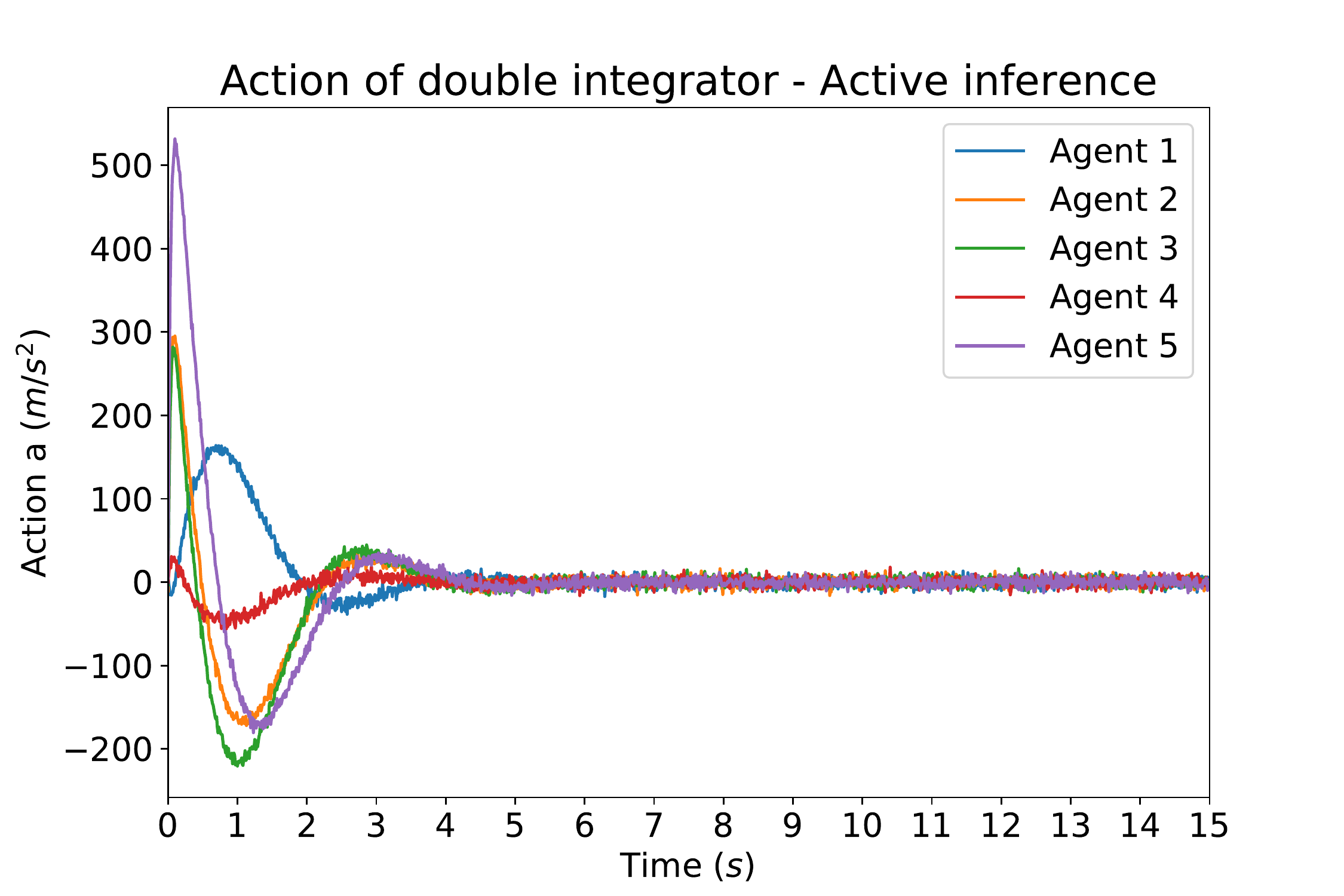}
		\caption{}
		\label{fig:DoubleIntegratorActiveInferenceAction}
	\end{subfigure}
  \caption{\textbf{The double integrator solved using active inference ($\bm{\alpha_1 = \exp (2), \alpha_2 = \exp (1)}$).} Same layout as \figref{fig:DoubleIntegratorLQG}. (a) Five examples with different initial conditions showing in blue the observed trajectories of different blocks in the phase-space and in red the agent's estimates of the same trajectories. (b) Actions taken by the five agents.}
  \label{fig:DoubleIntegratorActiveInference}
\end{figure}
\begin{figure}[ht!]
  \begin{subfigure}{.5\textwidth}
    \centering
    \includegraphics[width=\linewidth]{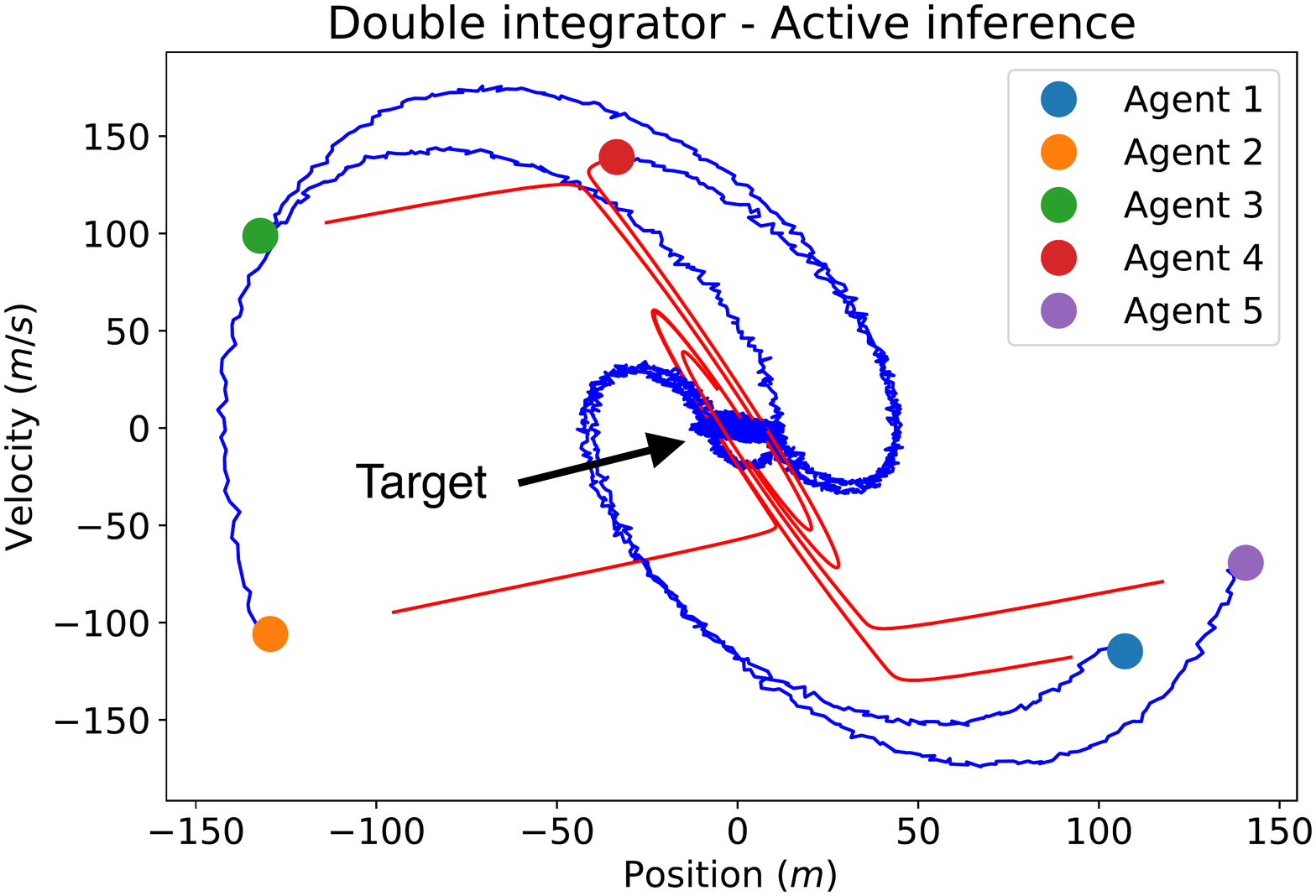}
    \caption{}
    \label{fig:DoubleIntegratorActiveInferenceNoExternalForceAgents}
  \end{subfigure}
  \begin{subfigure}{.5\textwidth}
    \centering
    \includegraphics[width=\linewidth]{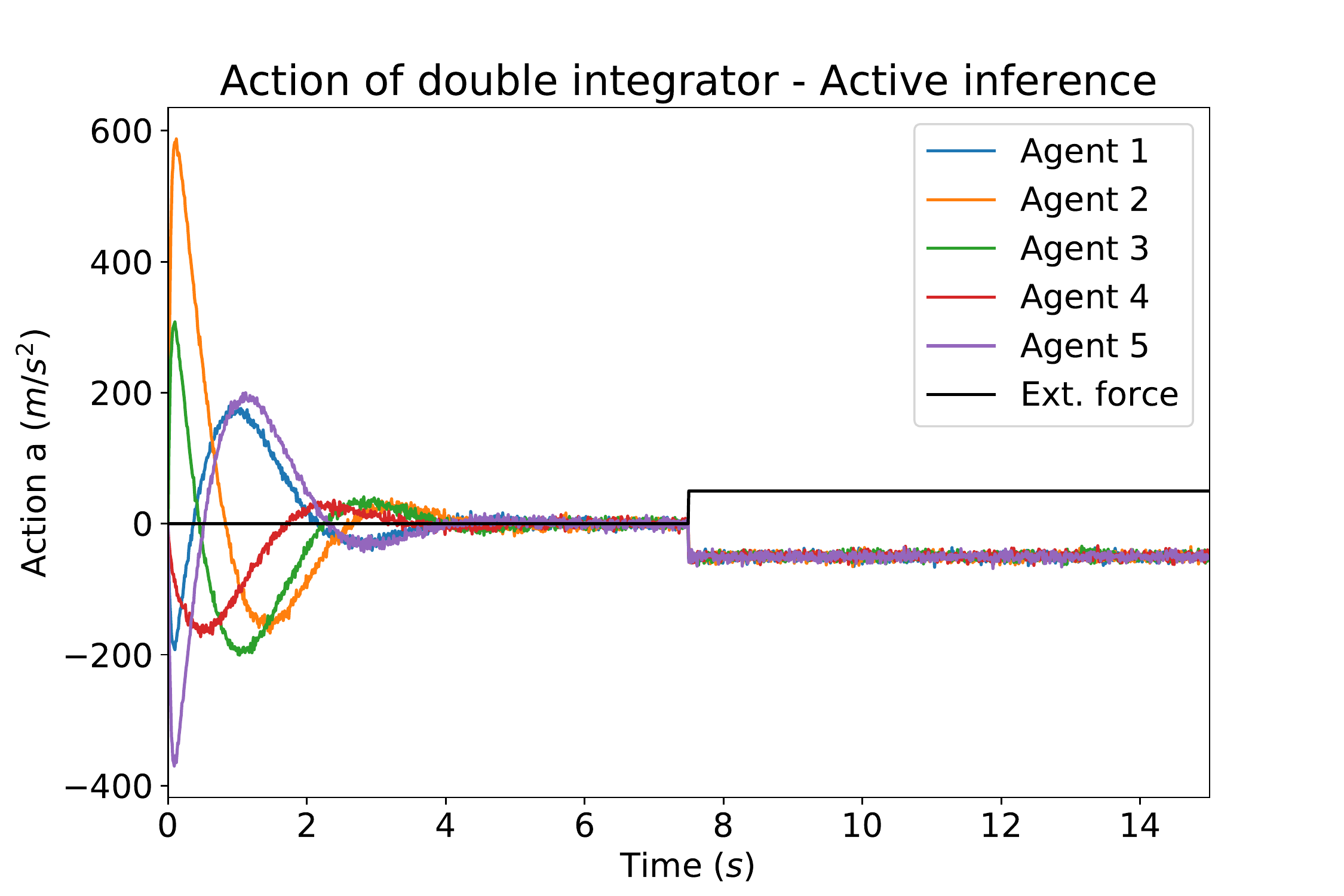}
    \caption{}
    \label{fig:DoubleIntegratorActiveInferenceNoExternalForceAction}
  \end{subfigure}
  \caption{\textbf{The double integrator solved using active inference ($\bm{\alpha_1 = \exp (2), \alpha_2 = \exp (1)}$).} Same layout as \figref{fig:DoubleIntegratorLQG}. (a) Five examples with different initial conditions showing in blue the observed trajectories of different blocks in the phase-space and in red the agent's estimates of the same trajectories. (b) Actions taken by the five agents after an external force is introduced (black line).}
  \label{fig:DoubleIntegratorActiveInferenceNoExternalForce}
\end{figure}
In \figref{fig:DoubleIntegratorActiveInference} we can see an example implementation of the double integrator using active inference. Five agents are initialised at random position and velocity (zero-mean Gaussian distributed, sd=300) and converge to the target solution where the output actions are essentially zero (excluding some noise), as expected \figref{fig:DoubleIntegratorActiveInferenceAction}. The most striking feature is that estimates of both position and velocity of the block are very inaccurate but the agent nonetheless reaches the desired target in the phase space, \figref{fig:DoubleIntegratorActiveInferenceAgents}. These differences are given by the generative implemented by the agent, encoding an imaginary spring-damper system that pulls it towards its desired state \figref{fig:DoubleIntegratorGM}. \figref{fig:DoubleIntegratorActiveInferenceNoExternalForce} shows the robustness of this implementation when an external force is introduced: by implementing integral control \cite{baltieri2018propabilistic}, active inference can in this case counteract the effects of unexpected inputs. The presence of integral action perfectly counteracts the effects of disturbances \figref{fig:DoubleIntegratorActiveInferenceNoExternalForceAction} (cf. \figref{fig:DoubleIntegratorLQGNoExternalForceAction}), and more importantly allows for the desired regulation of the agents' positions, \figref{fig:DoubleIntegratorActiveInferenceNoExternalForceAgents}, which is impossible in LQG accounts assuming perfect knowledge of the world (cf. \figref{fig:DoubleIntegratorLQGNoExternalForceAgents}).

\section{Discussion}
LQG-based architectures are modular in nature, with perception and action seen as separate problems solved nearly independently. According to this view, a system should initially find accurate estimates of the hidden properties of its observations, and only once such estimates are available should an agent attempt to regulate variables that are of interest to achieve its goals, e.g., temperature, oxygen level, etc.. On the other hand, we can define a framework based on mathematical formulations of control problems where the separation principle is not included or required. According to one such proposal, that we identified in active inference \cite{Friston2010biocyb, Friston2010nature}, perception and action are combined in an inseparable sensorimotor loop described by the minimisation of variational free energy for an agent. In this set up, action and perception are seen as instances of a fundamentally unique process \cite{clark1998being}, using different labels for our (i.e., the observers') convenience. In particular, the idea of precise inferences of world variables is called into question \cite{clark2015radical, baltieri2017active}, to the point that inaccurate perception is not only possible but becomes a pre-requisite to act on the world \cite{brown2013active, wiese2016action}. In architectures based on the separation principle, the estimated state of a system is thought of as a relevant account of real observations, e.g., their means and covariances. Conversely, in active inference it becomes clear that estimates of latent variables of the world are deeply connected to the current goal of an agent, e.g., to regulate its observations, cf. \cite{powers1973behavior}. To do so, its targets are encoded as prior expectations and used to bias inferential processes toward its desires so that prediction errors are created as the mismatch of observations and the estimates of hidden variables. These errors are then minimised by acting on the world \cite{Friston2010biocyb}, taking advantage of proprioceptive prediction errors that enact reflex arcs to make observation better accord with existing predictions \cite{clark2015surfing, wiese2016action}. More in general, the active inference formulation allows also for accurate estimates of the latent variables generating observations, see for instance \cite{Friston2008a}, but this modality fundamentally excludes the possibility of acting: if no prediction errors are generated for action to minimise, an agent becomes a simple mirror of its world with no strong desire or even necessity to act \cite{friston2012dark, brown2013active, baltieri2017active}. In other words, depending on different precision weights an agent can accurately estimate its observations without acting or potentially discard its sensations to only pursue its desires, generating all possible cases in between as a balanced mix of weighted prediction errors \cite{allen2018cognitivism}.

\section{Conclusions}
In recent years the more traditional understanding of perceptual and motor as nearly independent processes as been put into discussion by different authors, especially in neuroscience \cite{ahissar2016perception, busse2017sensation, buckley2018theory}. It is clear that many experimental set ups are limited \cite{krakauer2017neuroscience}, requiring new and ethologically meaningful paradigms for an appropriate study of different aspects of living systems \cite{najafi2018perceptual}. In this context, we propose some new ideas that could drive future experiments. These ideas are centred around a critical appraisal of LQG as a model architecture for cognitive systems, focusing in particular on the assumptions made by the use of Kalman-Bucy filters, central to these proposals \cite{todorov2002optimal, wolpert2011principles, franklin2011computational}. One of the key requirements for Kalman-Bucy filters to generate an accurate estimate of the hidden state of a system is to have access to \emph{all} the outputs (the observations) and \emph{all} the inputs (forces that affect the state) of a system. The inputs, in particular, include both motor commands, which in classical forward/inverse models are identified using the idea of efference copy \cite{von1950reafferenzprinzip} (see for instance \cite{kawato1999internal, wolpert2000computational, todorov2004optimality}), and external forces/signals from the environment that cannot be in principle accounted by an organism, i.e., a sudden change in weather conditions or unexpected interactions with other agents.

In this work we focused on the latter, since the presence of external unaccounted forces is often overlooked in many experimental set-ups with fixed or predictable conditions (e.g., the classic and still dominating two-alternative forced choice paradigm). In more realistic and ethological scenarios, however, one should expect that external and unpredictable stimuli constantly affect the behaviour of an agent \cite{krakauer2017neuroscience, najafi2018perceptual, buckley2018theory}. In this case, introducing noise or varying experimental conditions may help in testing the robustness of LQG-based architectures. In practice, if some inputs are not known, one should expect LQG to perform rather poorly until these inputs can be estimated and adaptation (e.g., learning) to new conditions can take place. However, one should then explain how such forces can be described in LQG since Kalman-Bucy filters cannot estimate inputs \cite{chen2003bayesian} (cf. DEM \cite{Friston2008a}). More in general, if a system is well adapted to deal with unpredictable stimuli, simple mechanisms such as integral control could be in place, as shown formally in \cite{sontag2003adaptation} and in experiments on chemotactic adaptation in E. Coli \cite{yi2000robust} for instance. More recently, some promising results have been presented in \cite{ritz2018control}, supporting the idea that integral feedback control, unlike Kalman(-Bucy) filters, is a good model for adaptation in environments with varying conditions. Integral control constitutes a linear approximation to problems of control with unknown forces affecting the observations of an agent \cite{aastrom2010feedback, baltieri2018propabilistic}, providing a robust solution with fast responses to problems that otherwise would require slower learning mechanisms \cite{yi2000robust}, which may be ineffective in fast-paced environments \cite{ashby1957introduction}.

\section{Acknowledgments}
This work was supported in part by a BBSRC Grant BB/P022197/1.

\footnotesize
\bibliographystyle{IEEEtran}


\end{document}